\documentclass [12pt]{article}
\pdfoutput=1
\usepackage{graphicx,amssymb,amsfonts,latexsym,amsmath,amsthm,times}
\usepackage{epsfig}
\usepackage{fancyhdr}
\usepackage{color}
\usepackage[english]{babel}
\numberwithin{equation}{section}

\begin{document}


\thispagestyle{plain}

\vspace*{2cm} \normalsize \centerline{\Large \bf Noise Induced
Phenomena in the Dynamics of Two Competing Species}

\vspace*{1cm}

\centerline{\bf D. Valenti$^a$\footnote{Corresponding author.
E-mail: davide.valenti@unipa.it}, A. Giuffrida$^b$, G. Denaro$^a$,
N. Pizzolato$^a$, L. Curcio$^a$, B.
Spagnolo$^{a,c,d}$,}\centerline{\bf S. Mazzola$^e$, G. Basilone$^e$,
A. Bonanno$^e$}

\vspace*{0.5cm}

\centerline{$^a$ Dipartimento di Fisica e Chimica, Universit\`a di
Palermo,}\centerline{Group of Interdisciplinary Theoretical Physics
and CNISM, Unit\`a di Palermo,}\centerline{Viale delle Scienze,
Ed.~18, I-90128~Palermo, Italy}
\medskip
\centerline{$^b$ Dipartimento di Scienze Veterinarie, Universit\`a
di Messina,}\centerline{Polo Universitario dell'Annunziata, 98168
Messina, Italy}
\medskip
\centerline{$^c$ Radiophysics Department, Lobachevsky State
University, Nizhniy Novgorod, Russia}
\medskip
\centerline{$^d$ Istituto Nazionale di Fisica Nucleare, Sezione di
Catania, Italy}
\medskip
\centerline{$^e$ Istituto per l'Ambiente Marino Costiero, CNR,
U.O.S. di Capo Granitola}\centerline{Via del Faro 3, I-91020
Campobello di Mazara (TP), Italy}


\vspace*{1cm}

\noindent {\bf Abstract.} Noise through its interaction with the
nonlinearity of the living systems can give rise to
counter-intuitive phenomena. In this paper we shortly review noise
induced effects in different ecosystems, in which two populations
compete for the same resources. We also present new results on
spatial patterns of two populations, while modeling real
distributions of anchovies and sardines. The transient dynamics of
these ecosystems are analyzed through generalized Lotka-Volterra
equations in the presence of multiplicative noise, which models the
interaction between the species and the environment. We find noise
induced phenomena such as quasi-deterministic oscillations,
stochastic resonance, noise delayed extinction, and noise induced
pattern formation. In addition, our theoretical results are
validated with experimental findings. Specifically the results,
obtained by a coupled map lattice model, well reproduce the spatial
distributions of anchovies and sardines, observed in a marine
ecosystem. Moreover, the experimental dynamical behavior of two
competing bacterial populations in a meat product and the
probability distribution at long times of one of them are well
reproduced by a stochastic microbial predictive model.
\vspace*{0.5cm}




\vspace*{1cm}

\setcounter{equation}{0}

\section{Introduction}\label{S:1}

\quad\enspace During last years, theoreticians worked to devise
deterministic mathematical models able to describe ecosystems in
which chaotic dynamics and spatial patterns are present. However,
due to their deterministic nature these models cannot reproduce or
explain the effects of random fluctuations, which come from the
intrinsic stochastic nature of open systems. Natural systems indeed
are a typical example of open systems due to the continuous presence
of deterministic and stochastic forces coming from the environment,
which affect the dynamics of these systems.

More recently, the role of noise in population dynamics has been the
subject of several theoretical studies~\cite{Bar93,Bjo01},
\cite{Cir03}-\cite{Ciu96},~\cite{Den13a,Den13b,Dro02,Dub08,Har14,Hie15,Hof08,LaB02,LaC10},~\cite{Roz99}-\cite{Ruo09a},~\cite{Sch01,Spa02a,Spa02b,Spa04a,Spa09,Sta01,Val12a,Vas07,Vil98,Zim99}.
Thus the study of the effects of noise is now a well established
subject in several different disciplines ranging from physics, to
chemistry and biology~\cite{Gat03,Leu01,Nau99,Per12,Val12b}.
However, the essential role of the noise in theoretical ecology has
been recently recognized. Some key questions in population ecology
are related to the comprehension of the role that noise, climatic
forcing and nonlinear interactions among individuals of the same or
different species play on the dynamics of the
ecosystems~\cite{Bjo01,Den13a,Den13b,Dro02,Val12a,Zim99}. Recently
researchers devoted more interest and attention to explain the role
of noise in different fields of
biology~\cite{Bar93,Bio99,Bro01,Chi05,Ciu93,Ciu96,Dub05,Fia08,Fre00,Gol99,Har14,Hig97,Hof08,Eco01,Tur00,Val08a,Vil98},
while investigating noise induced effects on population
dynamics~\cite{Cir03,LaB02,LaC10,Spa02a,Spa02b,Spa04a}.

In particular several works have studied the effects of random
fluctuations on the stability of ecological
systems~\cite{Bjo01,Zim99}, showing the presence of counterintuitive
phenomena, such as noise enhanced
stability~\cite{Agu01,Dub04,Man96,Spa04b,Spa07}, stochastic
resonance~\cite{Fia08,Gam98,Lan97,Man94,Man01}, and noise delayed
extinction~\cite{Cir03,LaB02,Spa02a,Spa02b,Spa03,Val04a}. The origin
of these effects is the interplay between nonlinear interactions
typical for natural systems and random fluctuations coming from
environment due to their intrinsic characteristic of open systems.
The permanent presence of noise and nonlinear interaction in
population dynamics causes indeed an increase of complexity compared
to other noise-driven systems, such as financial
markets~\cite{Bon06,Bon07,Spa08,Val07}, or many physical and
chemical processes described by deterministic dynamics~\cite{Eco01}.

Therefore the intrinsic nonlinearity can cause ecological systems to
critically depend on initial conditions, and both deterministic and
random perturbations coming from the environment. As a consequence,
the understanding of the role played by the noise in the dynamics of
nonlinear systems is a crucial point for a deeper comprehension and
successful modeling of the open systems governed by nonlinear
dynamics, which are known as complex systems.

Moreover we recall that the study of spatial distributions of
species densities is a major element to get a correct description in
population dynamics. An accurate analysis of spatio-temporal
patterns represents in fact a crucial point to devise predictive
models. As a consequence the comprehension of the role played by the
simultaneous presence of random fluctuations, deterministic forces
and nonlinear interaction, typical for natural systems, is
fundamental to effectively describe the spatio-temporal dynamics of
biological
populations~\cite{Bla99,Cir03,Kin01,LaB02,Rus00,Spa02a,Spa02b,Spa04a,Zho98}.
Thus, a deeper comprehension of the role of random fluctuations in
ecology underlies a better knowledge and description of real natural
systems. Nevertheless, despite of the big amount of theoretical work
and effort of researchers, a relevant biological issue such as the
role of environmental noise in ecological systems is still widely
debated.

For this purpose, we discuss here the effects of environmental noise
on the dynamics of biological populations. In particular, we review
some recent findings obtained for two competing species, whose
dynamics is described by generalized Lotka-Volterra equations, and
highlight the crucial role of the environmental noise on the
dynamics of the two species. Moreover we present new results (see
Section~\ref{S:3}), obtained by a coupled map lattice model, which
reproduce the spatial distributions of two fish populations, i.e.
anchovies and sardines. Our analysis focus on three different
ecosystems, in which external random fluctuations are modeled by
terms of multiplicative noise~\cite{Bar93,Ciu93,Ciu96}. As a result
we find that: (a) in a single compartment ecosystem the presence of
a driving force causes stochastic resonance, which results in
quasi-periodic oscillations of the two population densities; b) in
the same ecosystem it is possible to observe a nonmonotonic behavior
of the average extinction time of one species as a function of the
noise intensity, i.e. noise delayed extinction; (c) a
two-dimensional spatio-temporal model is able to reproduce the
concentrations of two fish species in a real marine ecosystem
located in the Mediterranean Sea; (d) in a food product the presence
of randomly fluctuating environmental variables such as temperature,
pH, and available water is taken into account in modeling the growth
of two bacterial populations, while allowing to obtain a better
agreement between experimental data and theoretical results compared
to the corresponding deterministic approach.

\section{Stochastic dynamics of two competing species: single compartment model}\label{S:2}

In this section we study the effect of random fluctuations in the
dynamics of two competing species. The model consists of generalized
Lotka-Volterra equations with terms of multiplicative noise, which
mimics the random fluctuations of environmental variables. The two
populations, $x$ and $y$, interact through the term $-\beta xy$,
where $\beta$ is the coupling constant, which regulates the
interaction strength and is a stochastic process. More in detail the
interaction coefficient $\beta(t)$ can be represented as a virtual
particle moving along a bistable potential and subject to both a
periodic driving term, which models seasonal changes in temperature,
and a term of additive noise which describes the effects due to the
noisy behaviour of the environment.

\subsection{The model}\label{S:2.1}

The stochastic dynamics of our ecosystem is given by the two
following generalized Lotka-Volterra equations~\cite{Lot20,Vol26}
\begin{eqnarray}
\frac{dx}{dt}=\mu_1\thinspace x\thinspace(\alpha_1-x-\beta_1(t) y)+x\thinspace\xi_x(t)\\
\frac{dy}{dt}=\mu_2\thinspace y\thinspace(\alpha_2-y-\beta_2(t)
x)+y\thinspace\xi_y(t),
 \label{LotVol}
\end{eqnarray}
which are stochastic differential equations in Ito sense, with
$\xi_x(t)$ and $\xi_y(t)$ statistically independent Gaussian white
noises with zero mean and correlation function $\langle
\xi_i(t)\xi_j(t')\rangle = \sigma \delta(t-t')\delta_{ij}$
($i,j=x,y$), and $\sigma$ the multiplicative noise intensity.
\begin{figure}[htbp]
\begin{center}
\includegraphics[width=12cm]{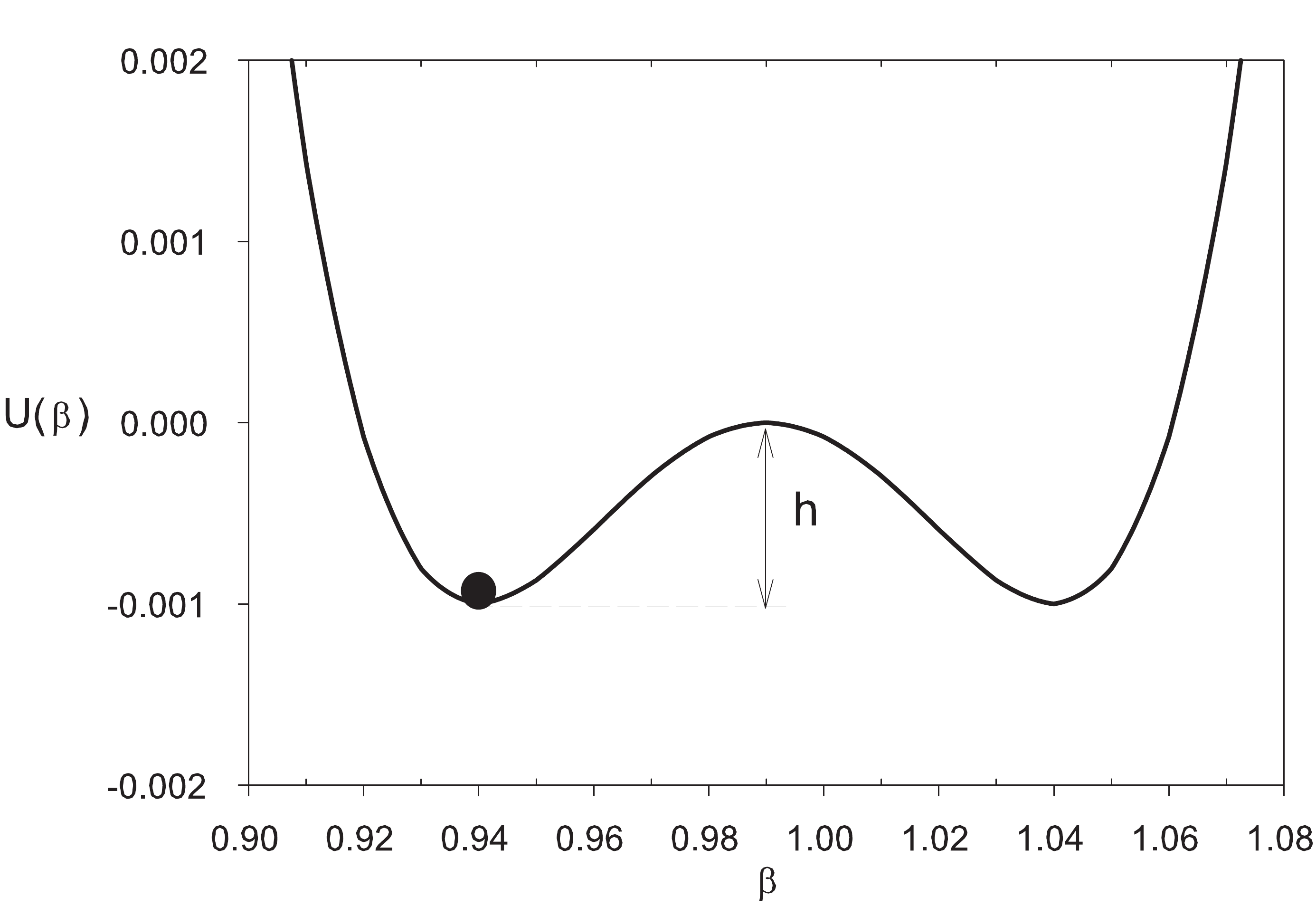}
\end{center}
\vspace{-0.8cm}\caption{ \small \emph{The bistable potential
$U(\beta)$ of the interaction parameter $\beta(t)$. The potential
$U(\beta)$ is centered on $\beta=0.99$. The parameters of the
potential are $h = 6.25 \cdot 10^{-3}$, $\eta=0.05$, $\rho =
-0.01$.}\bigskip} \label{potential} \vspace{-0.5cm}
\end{figure}
To get the time evolution for the two species densities the
parameters are set as follows: $\alpha_1=\alpha_2=\alpha $,
$\beta_1(t)=\beta_2(t)=\beta(t)$. We recall that $\beta < 1$
determines the coexistence regime (both species survives), while
$\beta> 1 $ corresponds to the exclusion regime (one of the two
species disappears after some time). Coexistence and exclusion of
one of the two species represent indeed stable states of the
Lotka-Volterra's deterministic model~\cite{Baz98}. As previously
said, natural systems are affected by a continuous exchange with a
noisy nonstationary environment. This implies that also the
interaction parameter is subject to both random fluctuations and
deterministic external signal such as the periodical changes in
temperature. The competition rate $\beta(t)$, continuously varying
between exclusion and coexistence regime due the interplay between
two main factors such as limiting resources and noisy environment,
is responsible for a random competition between the two populations.
As a consequence, the simultaneous presence of noise and periodic
driving causes the system to pass from a dynamical regime ($\beta <
1$, coexistence) to the other one ($\beta> 1$, exclusion) and vice
versa. This random process can be described by an Ito stochastic
differential equation, which reproduces, as previously noted, the
dynamics of a virtual particle moving along a bistable potential, in
the presence of a periodical driving force and an additive noise
term
\begin{equation}
\frac{d\beta(t)}{dt} = -\frac{dU(\beta)}{d\beta}+\gamma cos(\omega_0
t) + \xi_{\beta}(t) \label{beta_eq},
\end{equation}
where $U(\beta)$ is the bistable potential shown in Fig.~\ref{potential} and given by
\begin{equation}
U(\beta) = h(\beta-(1+\rho))^4/\eta^4-2h(\beta-(1+\rho))^2/\eta^2,
\label{U(beta)}
\end{equation}
Here $h$ is the height of the potential barrier. The periodic term
account for seasonal variations of environmental temperature, with
$\gamma=10^{-1}$ and $\omega_0/(2\pi)=10^{-3}$. In
Eq.~(\ref{beta_eq}) $\xi_{\beta}(t)$ is a Gaussian white noise with
the usual statistical properties $\langle \xi_{\beta}(t)\rangle=0$
and $\langle \xi_{\beta}(t)\xi_{\beta}(t')\rangle =
\sigma_{\beta}\delta(t-t')$, with $\sigma_{\beta}$ the additive
noise intensity. According to the form of the potential, one can
expect that in the deterministic case ($\sigma_\beta=0$) the
coexistence regime takes place when the virtual particle (initial
value of $\beta$) is placed in the left well.

\subsection{Stochastic resonance}\label{S:2.2}

As a first step we study how the noise affects the dynamics of the
two populations, whose dynamical regime is strongly dependent on the
interaction parameter. Therefore we analyze the time evolution of
$\beta(t)$ fixing different values of the additive noise intensity.
\begin{figure}[h!]
\begin{center}
\includegraphics[width=14cm]{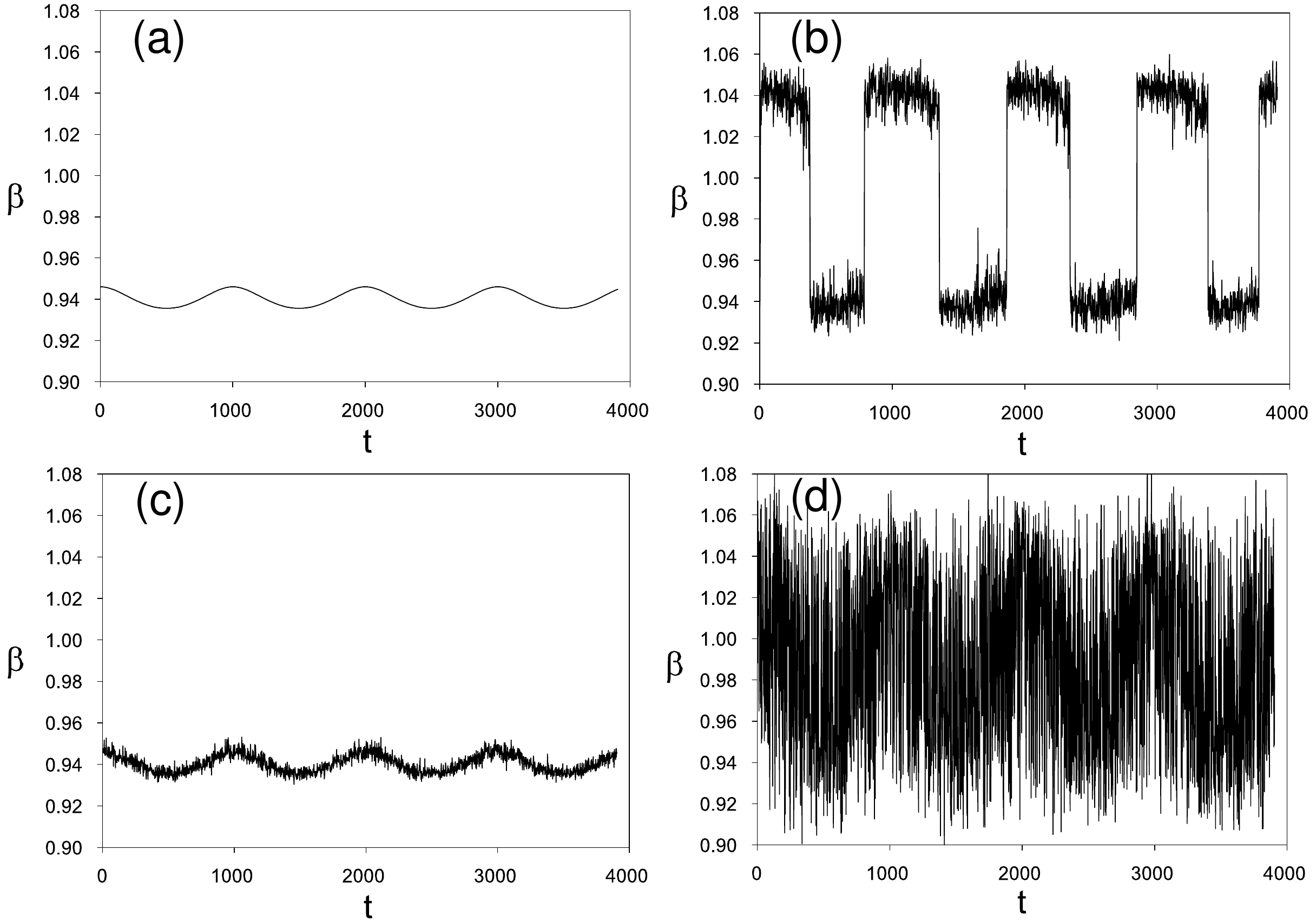}
\end{center}
\vspace{-0.8cm}\caption{ \small \emph{Time evolution of the
interaction parameter for different values of the additive noise
intensity. (a) $\sigma_\beta=0$; (b) $\sigma_\beta=1.78\cdot
10^{-4}$; (c) $\sigma_\beta=1.78\cdot 10^{-3}$; (d)
$\sigma_\beta=1.78\cdot 10^{-2}$. The values of the parameters are
$\gamma=10^{-1}$, $\omega_0/(2\pi)=10^{-3}$.}\bigskip}
\vspace{-0.5cm} \label{beta_series}
\end{figure}
\begin{figure}[h!]
\begin{center}
\includegraphics[width=14cm]{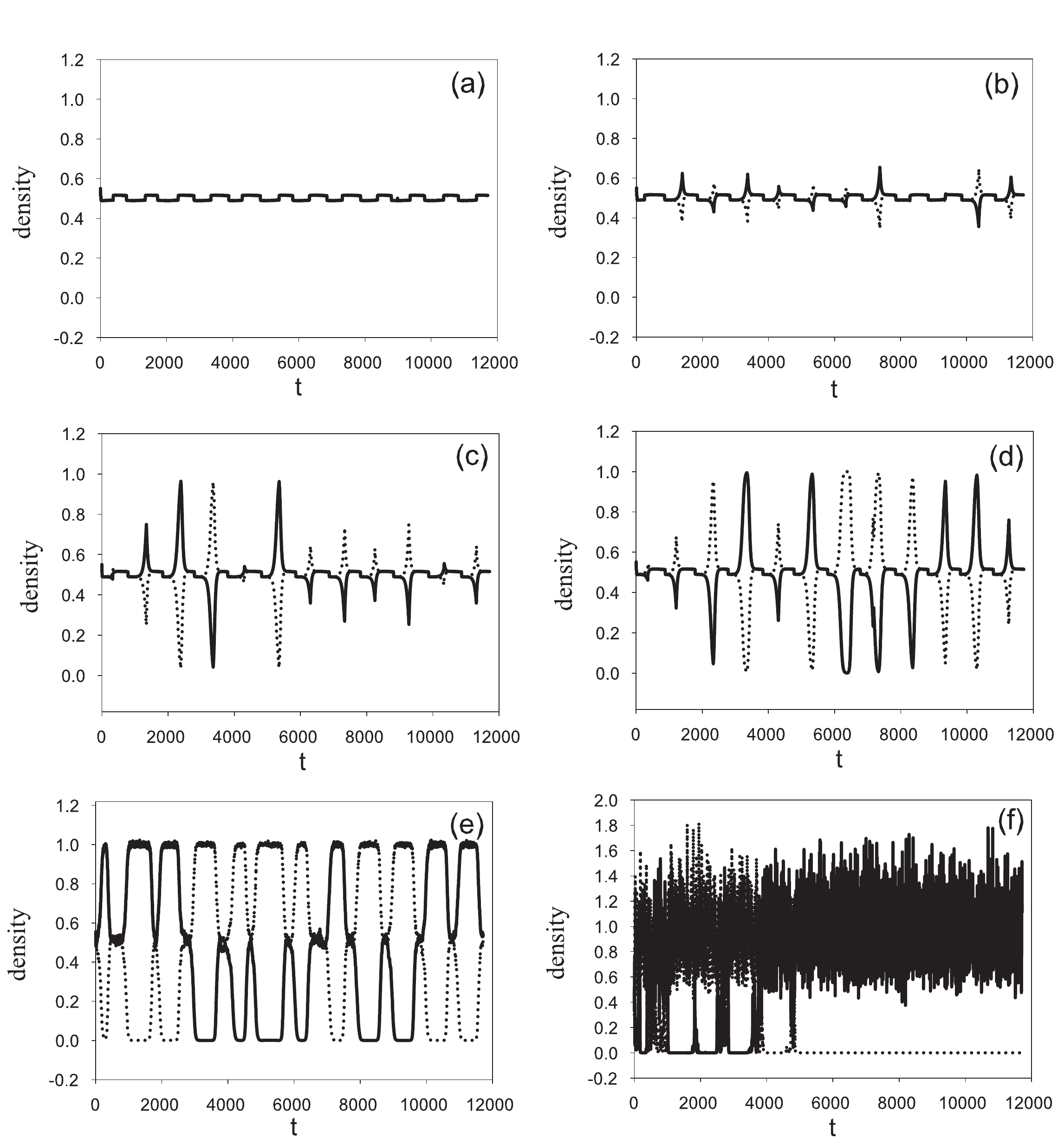}
\end{center}
\vspace{-0.8cm}\caption{ \small \emph{Time evolution of both
populations at different levels of the multiplicative noise: (a)
$\sigma=0$; (b) $\sigma=10^{-11}$; (c) $\sigma=10^{-10}$; (d)
$\sigma=10^{-9}$; (e) $\sigma=10^{-4}$; (f) $\sigma=10^{-1}$. The
values of the parameters are $\mu = 1$, $\alpha=1$, $\gamma =
10^{-1}$, $\omega_0/2\pi = 10^{-3}$. The intensity of the additive
noise is fixed at the value $\sigma_\beta=1.78 \cdot 10^{-3}$. The
initial values of the two species are $x(0)=y(0)=1$.}\bigskip}
\vspace{-0.5cm} \label{time_series}
\end{figure}
Setting $\sigma_\beta=0$ (deterministic regime), $\beta(t)$
undergoes a periodical behavior with the system remaining in the
coexistence regime (see Fig.~\ref{beta_series}a). Increasing the
noise ($\sigma_\beta \ll h$) the periodical behavior appears
slightly perturbed due to the presence of random fluctuations
(Fig.~\ref{beta_series}c). As the noise intensity increases
($\sigma_\beta \simeq h$), the virtual particle, i.e. the value of
the interaction parameter, jumps between $\beta=0.94$ and
$\beta=1.04$, which correspond to the coexistence ($\beta<1$) and
exclusion regime ($\beta>1$), respectively. In
Fig.~\ref{beta_series}b the typical picture of stochastic resonance
is shown. However, for higher values of the noise intensity
$\sigma_\beta$, a loss of coherence is observed and the system
dynamics is mainly driven by the source of random fluctuations (see
Fig.~\ref{beta_series}d). This behavior can be interpreted as a
cooperation between the periodical driving of the temperature, due
to some geological cause, and the environmental
noise~\cite{All01,Ben81,Ben82} for intermediate values of the noise
intensity. In this case the noise results to be tuned with the
deterministic oscillating external perturbation, causing the well
known stochastic resonance phenomenon. The noise intensity,
$\sigma_\beta=1.78 \cdot 10^{-3}$, which causes the synchronization
shown in Fig.~\ref{beta_series}c, can be easily obtained by the
formula~\cite{Jun89,Jun91} setting
\begin{equation}
\tau_k=T_0/2, \label{SR_sync}
\end{equation}
where $\tau_k$ is the Kramers time~\cite{Han90}
\begin{equation}
\tau_k = \frac{2 \pi}{\sqrt{\vert U''(0.99) \vert U''(0.94)}}
\exp{[2 h/\sigma_\beta]}, \label{kramers}
\end{equation}
and $T_0$ is the period of the driving force. In Eq.~(\ref{kramers})
$U''(0.99)$ and $U''(0.94)$ are the second derivative calculated in
the unstable and stable states of the potential, respectively.
Setting $\sigma_\beta=1.78\cdot 10^{-3}$, which gives an alternated
regime with quasi-periodical jumps between coexistence and
exclusion, it is possible to get the ecosystem dynamics when the
stochastic resonance (SR) condition is present, varying the
magnitude $\sigma$ of the multiplicative noise sources which act
directly on the two species. Using as initial conditions $x(0) =
y(0) = 1$, we note that, after a short transient, both populations
take on the same stationary value $x_{st} = y_{st} =
\alpha/(1+\beta) \approx 1/2$, around which the population densities
perform quasi-periodic oscillations whose amplitudes depend on the
magnitude of the multiplicative noise.

We observe that in the absence of multiplicative noise, i.e.
$\sigma=0$ (see Fig.~\ref{time_series}a), and for low noise
intensity, i.e. $\sigma=10^{-12}$ (see Fig.~\ref{time_series}b), a
dynamical coexistence regime characterized by correlated
oscillations of the two species densities is established. We observe
that this behaviour is connected with the symmetry of the
Lotka-Volterra equations due to the choice of the parameter values
and initial conditions (the same for the two populations). As a
consequence the species undergoes correlated oscillations around the
stationary value $\alpha/(1+\beta)$ (see panels a, b of
Fig.~\ref{time_series}), even if the ecosystem, driven by the
quasi-periodic behaviour of the interaction parameter (SR effect),
is in the exclusion regime during the $50\%$ of the time.

The symmetry condition is broken as the multiplicative noise
intensity increases, causing the appearance of anti-correlated
oscillations (see panels c, d, e of Fig.~\ref{time_series}). The
amplitude of these anti-correlated oscillations increases for higher
intensity of the multiplicative noise ($\sigma=10^{-2}$), causing a
degradation of the quasi-periodical behaviour in the time series of
the populations (see Fig.~\ref{time_series}f): now the ecosystem is
mainly driven by the noise which tends to suppress the effect of the
periodical signal. The presence of the multiplicative noise indeed
breaks the symmetric dynamical behaviour of the ecosystem. For
$\beta>1$, i. e. in the exclusion regime, this symmetry breaking
determines different behaviour in the time series of the two
species: one tends to survive, while the other one tends to
extinguish. Fig.~\ref{time_series} indicates that the effect of
quasi-periodic signal can be amplified by higher intensities of the
multiplicative noise. The periodicity of the noise-induced
oscillations in the time behavior of the population densities shown
in Fig.~\ref{time_series}e is the same of the driving periodic term
of Eq.(\ref{beta_eq}).
\begin{figure}[h!]
\begin{center}
\includegraphics[width=10cm]{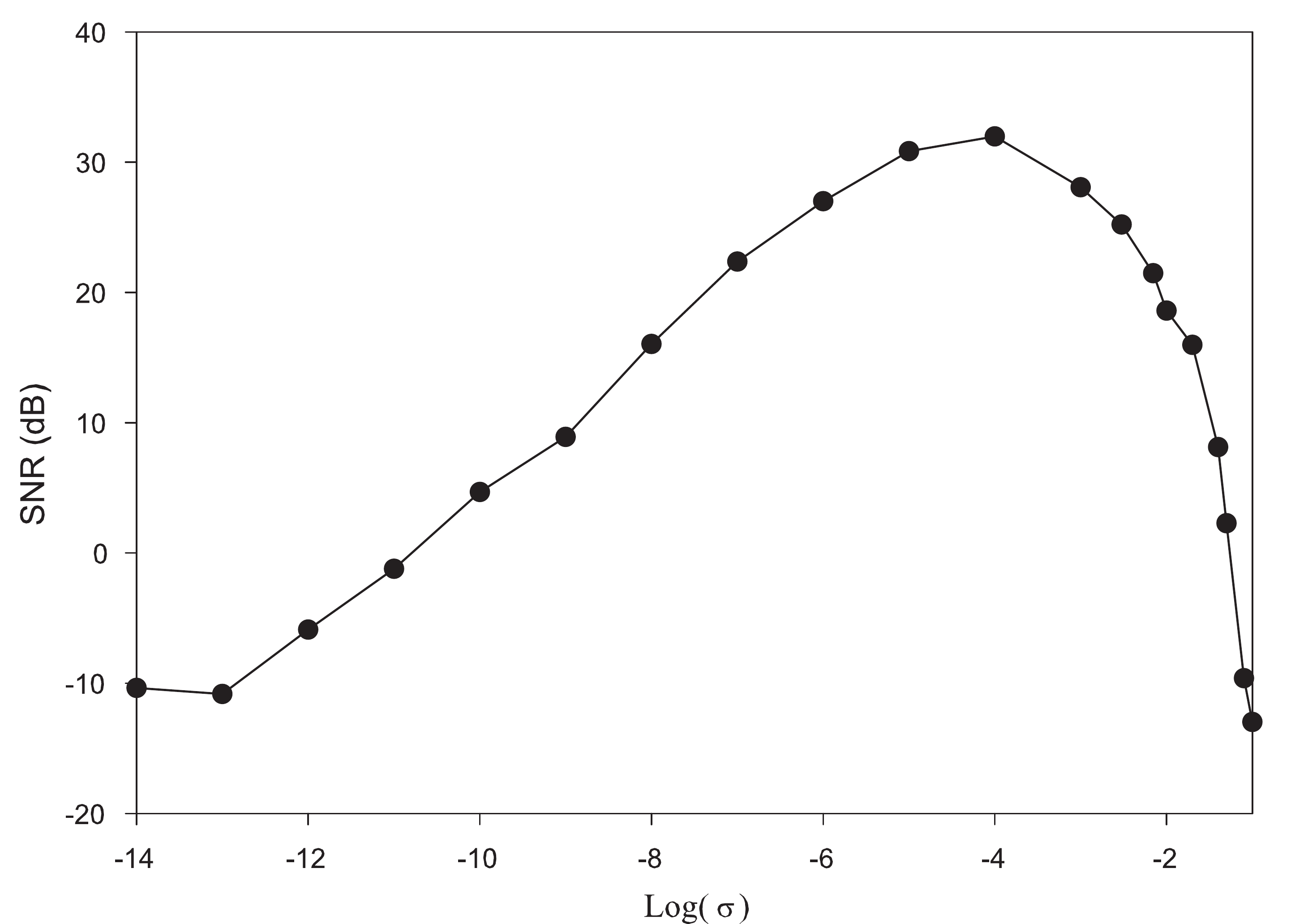}
\end{center}
\vspace{-0.8cm}\caption{ \small \emph{Log-Log plot of SNR as a
function of the multiplicative noise intensity. The SNR corresponds
to the squared difference of population densities $(x - y)^2$. The
values of the parameters are the same of
Fig.~\ref{time_series}.}\bigskip}\vspace{-0.5cm} \label{snr}
\end{figure}
This is the signature of stochastic resonance phenomenon. This
second SR phenomenon can be quantitatively investigated calculating
the signal-to-noise ratio (SNR) of the squared difference of
population densities. More exactly we consider the time series
$[x(t)~-~y(t)]^2$ at different values of the noise intensity
$\sigma$, and calculate the SNR according to~Ref.~\cite{Agu10}. This
quantity is shown in Fig.~\ref{snr} as a function of the
multiplicative noise intensity $\sigma$, for $\sigma_\beta=1.78
\cdot 10^{-3}$. We note that a maximum is found for
$\sigma=10^{-4}$. This indicates that the additive noise determines
the conditions for the different dynamical regimes of the two
species, while the multiplicative noise is responsible for a
coherent response of the system, breaking the initial condition of
symmetry of the ecosystem.

\subsection{Noise delayed extinction}\label{S:2.3}

In this section we study how one of the two populations can vanish
due to the interplay between the periodical signal and the additive
noise $\xi_{\beta}(t)$, which drive the ecosystem through the two
different regimes, i.e. coexistence and exclusion.

For this purpose we introduce the average extinction time of one
species and calculate it for different values of the noise intensity
$\sigma_\beta $, setting the multiplicative noise intensity at a
small value, so that the dynamics is weakly perturbed by the noise
and the ecosystem remains far from the SR regime.
\begin{figure}[h!]
\begin{center}
\vspace{0.0cm}\includegraphics[width=16cm]{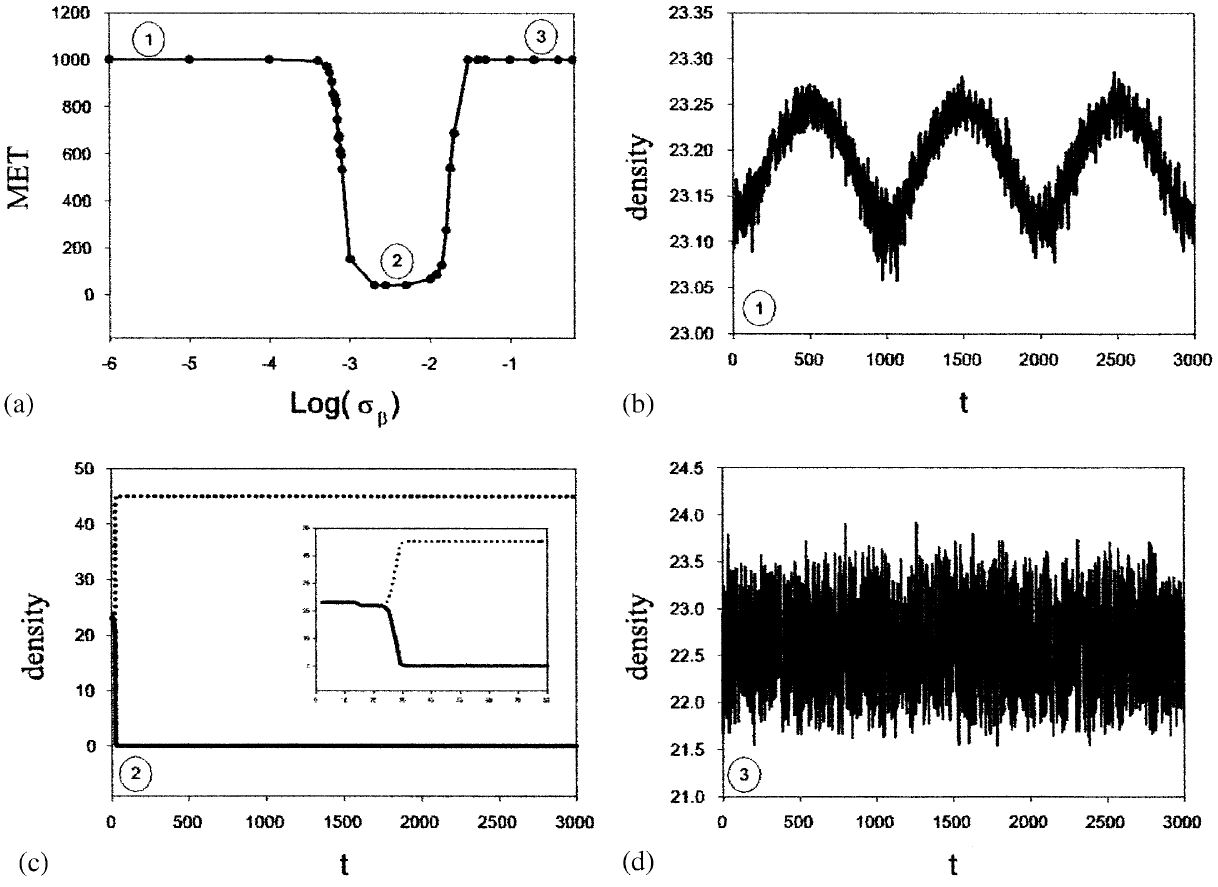}
\end{center}
\vspace{-0.8cm}\caption{\small \emph{(a) Mean extinction time of one
species as a function of the noise intensity $\sigma_\beta$. Time
evolution of both species for different levels of additive noise:
(b) $\sigma_\beta=10^{-4}$, (c) $\sigma_\beta=2 \cdot 10^{-3}$, (d)
$\sigma_\beta=10^{-1}$. The values of the parameters are $\mu = 1$,
$\alpha=45$, $\gamma = 10^{-1}$, $\omega_0/2\pi = 10^{-3}$. The
intensity of the multiplicative noise is fixed at the value
$\sigma=10^{-9}$. The initial values of the two species are
$x(0)=y(0)=1$.}\vspace{-0.5cm}\bigskip} \label{met}
\end{figure}
Fixing $\sigma=10^{-9}$ and $\beta(0)=0.94$ as initial condition, we
integrate Eqs.~(1)~and~(2) by performing 200 numerical realizations,
and obtaining the behaviour of the mean extinction time (MET) of one
species as a function of the additive noise intensity
$\sigma_{\beta}$ (see Fig.~(\ref{met}a)).

In these conditions the ecosystem is in the coexistence regime, that
is the deterministic extinction time of both species is infinite. By
introducing noise, exclusion takes place and a finite mean
extinction time (MET) appears.

We note that for small and large values of the additive noise
intensity (see regions 1 and 3 in Fig.~\ref{met}a), the system is in
a dynamical regime which favours the coexistence of the two species,
so that no extinction occurs (see time series shown in panels b and
d of Fig.~\ref{met}). Conversely, for intermediate values of
$\sigma_\beta$, exclusion of one species is found with different
extinction times. In particular, a minimum MET is observed for
$\sigma_\beta=2.75 \cdot 10^{-3}$, which is of the same order of
magnitude of the potential barrier height $h$. The Kramers time
corresponding to this noise intensity is $\tau_k=41.6$, a value
approximately equal to $\tau_{min}$. This behaviour reproduces the
effect of random modifications of the environmental conditions
responsible for the delayed extinction of biological species in real
ecosystems, as it is observed in experimental data of populations in
a very long time interval~\cite{Car02,Gar02}. Starting from a
situation in which one of the two population undergoes rapidly
extinction (region 2 in Fig.~\ref{met}a), varying the noise
intensity (decrease or increase of $\sigma_\beta$) the ecosystem
moves towards regions where the MET becomes larger (see the region
between 1 and 2 and that between 2 and 3 in Fig.~\ref{met}a). From a
physical point of view this effect can be explained observing that
for small values of $\sigma_{\beta}$, the initial condition
($\beta(0)=0.94$) is maintained for low noise intensity (the Kramers
time is very long and the virtual particle takes very long times to
cross the barrier and reach the right well corresponding to the
exclusion regime). On the other side, for higher noise intensities
there is a strong decrease of the Kramers time and the values of
$\beta(t)$ switch very rapidly between the two wells, determining an
alternate regime coexistence/exclusion which avoids the extinction.
Conversely, for intermediate values of the additive noise intensity
(see region 2 in Fig.~\ref{met}), after an initial permanence in the
left well $\beta(t)$ reaches the right well, where it remains for a
time long enough to cause the extinction of one of the two species
(see time series shown in panel c of Fig.~\ref{met}).

\section{Stochastic dynamics of two competing species: spatially extended model}\label{S:3}

In this section we consider two populations, distributed in a
two-dimensional spatial domain, subject to multiplicative noise, in
the presence of an external periodic signal. The multiplicative
noise source mimics the effects of random fluctuations of the
environmental variables.

\subsection{The model}\label{S:3.1}

The spatio-temporal dynamics of the two populations is given by a
discrete time evolution model, based on the coupled map lattice
(CML) approach~\cite{Kan92}, and is the discrete version of the
Lotka-Volterra equations, where diffusive terms were
added~\cite{Val04b}

\begin{eqnarray}
x_{i,j}^{n+1}&=&\mu x_{i,j}^n (1-x_{i,j}^n-\beta^n
y_{i,j}^n)+\sqrt{\sigma_x}
x_{i,j}^n X_{i,j}^n + D\sum_\gamma (x_{\gamma}^n-x_{i,j}^n) \label{Lot_one}\\
y_{i,j}^{n+1}&=&\mu y_{i,j}^n (1-y_{i,j}^n-\beta^n
x_{i,j}^n)+\sqrt{\sigma_y} y_{i,j}^n Y_{i,j}^n + D\sum_\gamma
(y_{\gamma}^n-y_{i,j}^n), \label{Lot_two}
\end{eqnarray}
with $x_{i,j}^n$ and $y_{i,j}^n$ densities of the two populations in
the site \textit{(i,j)} at the time step \textit{n}. Here $\beta^n$
is the interaction parameter at the same time step, $\mu$ is the
growth rate, \textit{D} is the diffusion constant and $\sum_\gamma$
represents the sum over the four nearest neighbors. $X_{i,j}^n$ and
$Y_{i,j}^n$ are independent Gaussian random variables with zero mean
and variance unit. Moreover, $\sigma_x$ and $\sigma_y$ are the
intensities of the two multiplicative noise sources. The interaction
parameter $\beta^n $ is the same stochastic process $\beta(t)$ as in
previous section (see Eq.~(\ref{beta_eq})), where $U(\beta)$ is the
bistable potential shown in Fig.~\ref{potential}. Therefore, also in
this spatially extended model the switching of $\beta(t)$ between
the exclusion and coexistence regime occurs randomly. This mimics
the effect of the noisy environment on limiting factors such as food
resources, with the periodical force accounting for the periodical
(seasonal) oscillations of temperature. According to the analysis
performed for the single compartment system, the two stable states
(left and right wells of the potential $U(\beta)$) correspond to
coexistence and exclusion of one of the two species of the
Lotka-Volterra deterministic
model~\cite{Cir03,LaB02,Spa02a,Spa02b,Spa04a,Val04a}.

\subsection{Results}\label{S:3.2}

\begin{figure}[h!]
\begin{center}
\includegraphics[width=5.5cm]{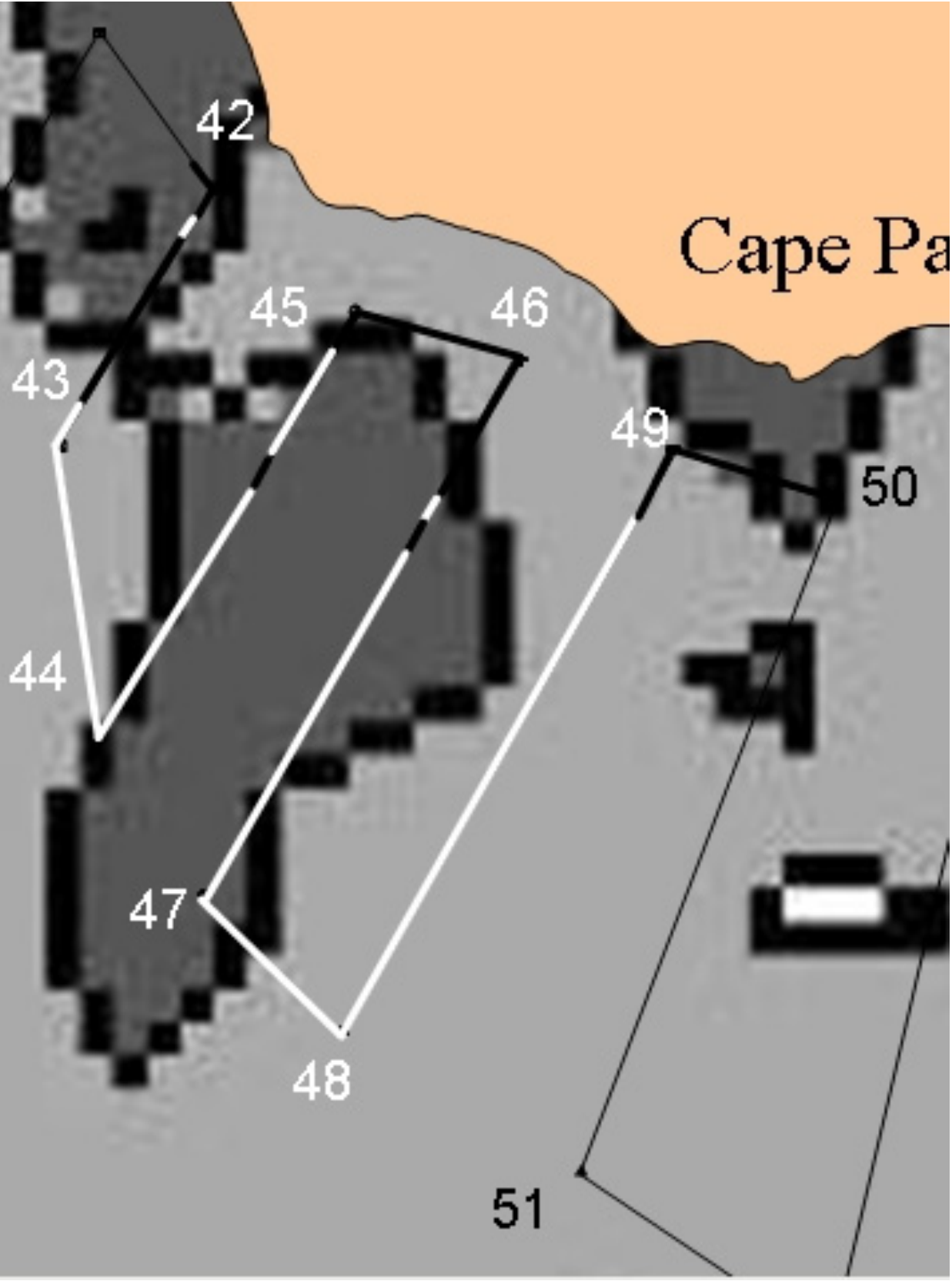}
\includegraphics[width=5.75cm]{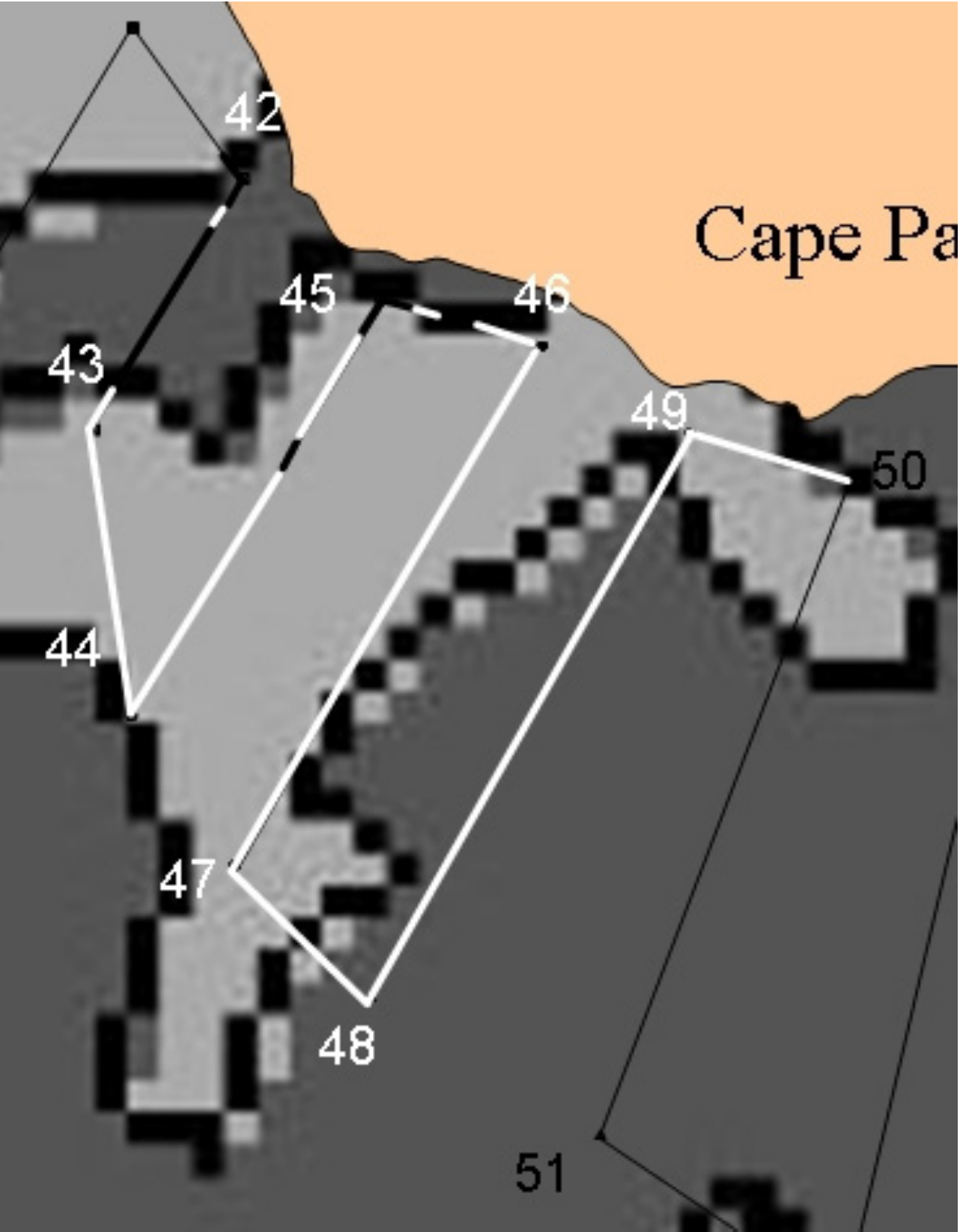}
\end{center}
\caption{\small \emph{Comparison between spatial distribution of
species $x$ and anchovy abundance (left panel) and species $y$ and
sardine abundance (right panel). The values for $x$ and $y$ were
obtained from the model at time step $n=600$. The anchovy and
sardine abundances were estimated experimentally along the acoustic
transect, from point 42 to point 50, tracked during the
oceanographic campaign "ANCHEVA '02" by the Interdisciplinary Group
of Oceanography of IAMC-CNR of Mazara del Vallo. The spatial
distribution of $x$ and $y$ are drawn by light and dark grey zones
which represent, respectively, low and high density of the two
species. The values of the parameters are $\mu = 2$, $\nu =
\omega_0/(2\pi)=0.34$, $\gamma=10^{-5}$,
$\sigma_x=\sigma_y=10^{-8}$, $\sigma_\beta=10^{-12}$, $D=0.05$,
$\beta(0)=0.95$. The initial values for the spatial distributions of
$x$ and $y$ are $x^{init}_{i,j}=y^{init}_{i,j}=0.5$ for all sites
$(i,j)$. Concerning the experimental distributions, white and black
tracts indicate, respectively, small and large values of anchovy
(left panel) and sardine (right panel) abundances estimated
experimentally during the oceanographic campaign~\cite{Bon02}.}}
\vspace{-0.2cm} \label{anchsard}
\end{figure}

By numerical integration of Eqs.~(\ref{Lot_one})-(\ref{beta_eq}) we
obtain the spatio-temporal evolution of the population densities for
the two theoretical species. The results are given in
Fig.~\ref{anchsard}.
Here the spatial distribution of the species $x$ and $y$ are shown
at a certain time in the left and right panel, respectively. Light
and dark grey zones represent low and high densities, respectively.
These theoretical spatial distribution are compared with real data,
for anchovies and sardines, collected along the acoustic transects
tracked by the Interdisciplinary Group of Oceanography of IAMC-CNR
of Mazara del Vallo~\cite{Bon02}, during the oceanographic campaign
"ANCHEVA '02" in the Strait of Sicily. In particular, in
Fig.~\ref{anchsard} (left panel) the anchovy abundance, estimated
experimentally along the acoustic transects, from point 42 to point
50 is shown. Analogously in Fig.~\ref{anchsard} (right panel) the
sardine abundance, estimated experimentally along the same acoustic
transects, is shown. White and black tracts indicate small and large
values, respectively, of anchovy and sardine abundances.\\
Predicted results show a good qualitative agreement with the spatial
distributions observed for anchovy and sardine abundances. In
particular we find that, in the most of the transects considered, a
higher (lower) density for species $x$ corresponds, in the same
area, to a larger (smaller) anchovy abundance. More in detail, in
Fig.~\ref{anchsard} (left panel) a good correspondence can be
observed between the density of the species $x$ and the anchovy
abundance along the segments 42-43, 43-44, 45-46, 47-48, 48-49 and
49-50, with discrepancies appearing along the segments
44-45 and 46-47.\\
A similar situation is observed for species $y$ and sardines. In
particular, the spatial distributions of species $y$ and sardines
are shown in Fig.~\ref{anchsard} (right panel). Here a good
agreement between predicted and observed abundances can be observed
along the segments 42-43, 43-44, 44-45, 45-46, 46-47, 47-48 and
49-50, with a lack of agreement along the segment 48-49.
\section{Predictive microbiology}
\label{S:4}
Predictive microbiology exploits mathematical models to describe
bacterial dynamics in different products of food industry. The
models take into account the role played by environmental variables,
whose variations can affect, sometimes dramatically, the quality and
safety of the food products. Predictive models belong to three
different types: primary, secondary and tertiary~\cite{Whi93}. The
first class of models allows to obtain the time evolution of
microbial populations. The models belonging to the second type give
information on the relationship between parameters which appear in
primary models, and physical and chemical variables such as T
(temperature), pH (hydrogen ion concentration), and aw (activity
water). The third class of models puts together the primary and
secondary ones, letting the evolution of physical and chemical
variables be considered, when analysis and prediction of the
concentration
of spoiling or pathogen bacteria of the food are performed~\cite{Dal02}.\\
\indent A well-known method for the theoretical analysis of
microbial growth exploits generalized Lotka-Volterra (LV)
equations~\cite{Lot20,Vol26}, which allow to describe the dynamics
of two competing bacterial populations in different food products. A
prototype model structure for mixed microbial populations in food
products was proposed by Dens et al.~\cite{Den99}. A similar
approach indicated that experimental data for \emph{Escherichia
coli} O157:H7 in ground beef could be well reproduced by an
interspecific competition model for two bacterial populations. In
the same work the effects of random fluctuations were considered
using growth rates whose values are obtained from uniform random
distributions~\cite{Pow04}. An extensive review on predictive
microbiology showed that in general a stochastic approach provides
predictions which exclude the worst-case scenario~\cite{McM02}. In
particular, stochastic terms were introduced to reproduce and
predict bacterial dynamics, exploiting an approach based on primary
and secondary growth models~\cite{Nau00}. Moreover other authors
presented a stochastic model which interprets the bacterial growth
as the average evolution of many cells: measured values of the
growth rate for many different cells allow to describe the
theoretical growth rate used in the model as a stochastic variable
with a corresponding probability
distribution~\cite{Bar98,Bar01,Swi04}.\\
\indent The previous models however do not include explicitly
stochastic terms in the equations of motion of the systems analyzed.
In other words, the models used in predictive microbiology
are not usually based on stochastic differential equations.\\
\indent In the following we analyze how predictions for bacterial
dynamics are affected by the three following features: (i) use of
differential equations (dynamical approach); (ii) presence of
interactions among bacterial populations; (iii) introduction of
stochastic terms, i.e. noise sources, which mimic the random
fluctuations of environmental variables.

\subsection{Bacterial growth in meat products: single compartment dynamics of two interacting populations}
\label{S:4.1}

In this section we introduce a model for the dynamics of two
competing bacterial populations, \emph{Listeria monocytogenes} and
lactic acid bacteria (LAB), present in a meat product, i.e. a
traditional Sicilian salami (Salame S. Angelo PGI (Protected
Geographical Indication)) very important from the point of view of
the Italian food industry. Specifically, \emph{L. monocytogenes} is
a microbial agent of foodborne disease, while LAB constitute the
normal bacterial flora of the substrate. The theoretical approach is
based on generalized Lotka-Volterra (LV)
equations~\cite{Den99,Pow04}, in which the bacterial growth rates
depend on environmental variables, such as temperature, pH, and
activity water, whose randomly fluctuating behaviour can be modeled
by inserting terms of additive white Gaussian noise:
\begin{eqnarray}
\frac{dN_{Lmo}}{dt}&=&\mu^{max}_{Lmo}\thinspace N_{Lmo}
\frac{Q_{Lmo}}{1+Q_{Lmo}}\left(1-\frac{N_{Lmo}+\beta_{Lmo/LAB}\thinspace
N_{LAB}}{N^{max}_{Lmo}}\right)\label{eq_Lmo}\\
\frac{dQ_{Lmo}}{dt}&=&\mu^{max}_{Lmo} \thinspace Q_{Lmo}\label{QLmo}\\
\frac{dN_{LAB}}{dt}&=&\mu^{max}_{LAB} \thinspace N_{LAB}
\frac{Q_{LAB}}{1+Q_{LAB}}\left(1-\frac{N_{LAB}+\beta_{LAB/Lmo}\thinspace
N_{Lmo}}{N^{max}_{LAB}}\right)\label{eq_LAB}\\
\frac{dQ_{LAB}}{dt}&=&\mu^{max}_{LAB}\thinspace Q_{LAB}.\label{QLAB}
\end{eqnarray}
Here, $N_{Lmo}$ and $N_{LAB}$ are the population concentrations of
\emph{L. monocytogenes} and LAB, respectively; $\mu_{Lmo}$ and
$\mu_{LAB}$ represent the maximum specific growth rates of the two
populations, and $N^{max}_{Lmo}$ and $N^{max}_{LAB}$ are the
theoretical maximum population concentrations. The coefficients
$\beta_{Lmo/LAB}$ and $\beta_{LAB/Lmo}$ are the interspecific
competition parameters of LAB on \emph{L. monocytogenes} and
vice-versa. $Q_{Lmo}$ and $Q_{LAB}$ represent the physiological
state of the two populations.

To solve Eqs.~(\ref{eq_Lmo})--(\ref{QLAB}) it is necessary to set
how $\mu^{max}_{Lmo}$ and $\mu^{max}_{LAB}$ vary. This can be done
by introducing for the maximum growth rates the following secondary
model
\begin{eqnarray}
\mu^{max}_{Lmo}&=&0.14776 \thinspace(T_0-0.88)
\cdot(1-\exp(0.536\thinspace(T-41.4))) \cdot\sqrt{aw-0.923}\nonumber\\
&\cdot&\sqrt{1-10^{4.97-pH}}\cdot\sqrt{1-\frac{LAC}{3.79\thinspace(1+10^{pH-3.86})}10^{4.97-pH}}\cdot\frac{350-NIT}{350}\label{muLmo}\qquad\\
\mu^{max}_{LAB}&=&0.00234\thinspace(aw-0.928)\cdot(pH-4.24)\cdot(pH-9.53)\cdot(T-3.63).\label{muLAB}\qquad
\end{eqnarray}
obtained by a phenomenological approach~(see Ref.~\cite{Giu09} and
references therein). Here, NIT is nitrite concentration in $ppm$ and
LAC is lactic acid concentration in $g l^{-1}$. The values 0.88,
41.4, 0.923, 4.97, and 350 represent $T_{min}$ ($^\circ C$),
$T_{max}$ ($^\circ C$), $aw_{min}$, $pH_{min}$ and $NIT_{max}$,
respectively. Temperature, pH, and activity water are described as
stochastic processes. In particular, their dynamics is given by two
different contributions: (i) a linearly decreasing deterministic
behaviour within a time interval of 168 h, according to the
procedure followed in the production process (a fermentation period
of 7 days); (ii) terms of additive white Gaussian noise, which
account for the presence of random fluctuations due to environmental
perturbations. By this way the following system of three stochastic
differential equations is obtained~\cite{Giu09}
\begin{eqnarray}
\frac{d\thinspace T(t)}{dt}&=&k_T\thinspace t + \xi_T(t)\label{eqT}\\
\frac{d\thinspace pH(t)}{dt}&=&k_{pH}\thinspace t + \xi_{pH}(t)\label{eqpH}\\
\frac{d\thinspace aw(t)}{dt}&=&k_{aw}\thinspace t +
\xi_{aw}(t),\label{eqaw}
\end{eqnarray}
where $\xi_i(t)$, with $i\thinspace=\thinspace T, pH, aw$, are
statistically independent Gaussian white noises with the following
properties
\begin{eqnarray}
&&<\xi_i(t)>=0\\
&&<\xi_i(t)\xi_i(t')>\thinspace=\thinspace\sigma_i \delta(t-t'),
\label{noise}
\end{eqnarray}
and $\sigma_i$ are the noise intensities.
\begin{figure}[h!]
\begin{center}
\includegraphics[width=12cm]{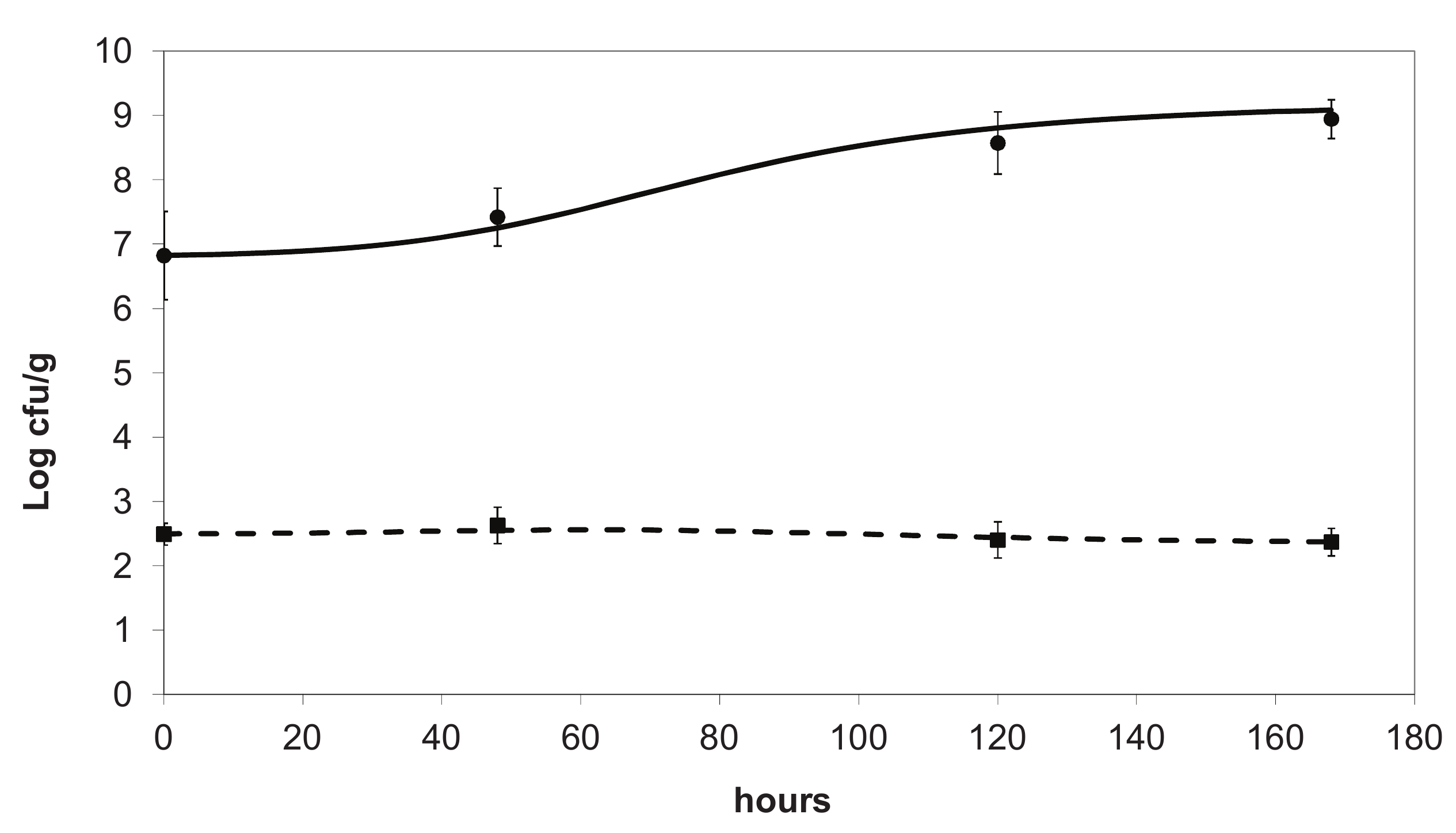}
\end{center} \vspace{-0.8cm}\caption{Theoretical growth curves for \emph{L.
monocytogenes} (dashed black line) and LAB (full black line), and
corresponding experimental data (black squares for \emph{L.
monocytogenes}, black circles for LAB). Vertical bars indicate the
experimental errors.}\label{fig7}
\end{figure}
Eqs.~(\ref{eq_Lmo})--(\ref{eqaw}) have been solved numerically
within the Ito scheme, performing 1000 realizations and obtaining
the mean growth curves in absence of noise ($\sigma_T=0$,
$\sigma_{pH}=0$, $\sigma_{aw}=0$). The initial concentrations of the
two populations, however, have been set randomly. Specifically, in
each realization the initial values of $N_{Lmo}$ and $N_{LAB}$ have
been extracted from two Gaussian distributions, whose mean values
and standard deviations were equal to those of the distributions
experimentally observed~\cite{Giu05}.
\begin{figure}[h!]
\begin{center}
\includegraphics[width=13.0cm]{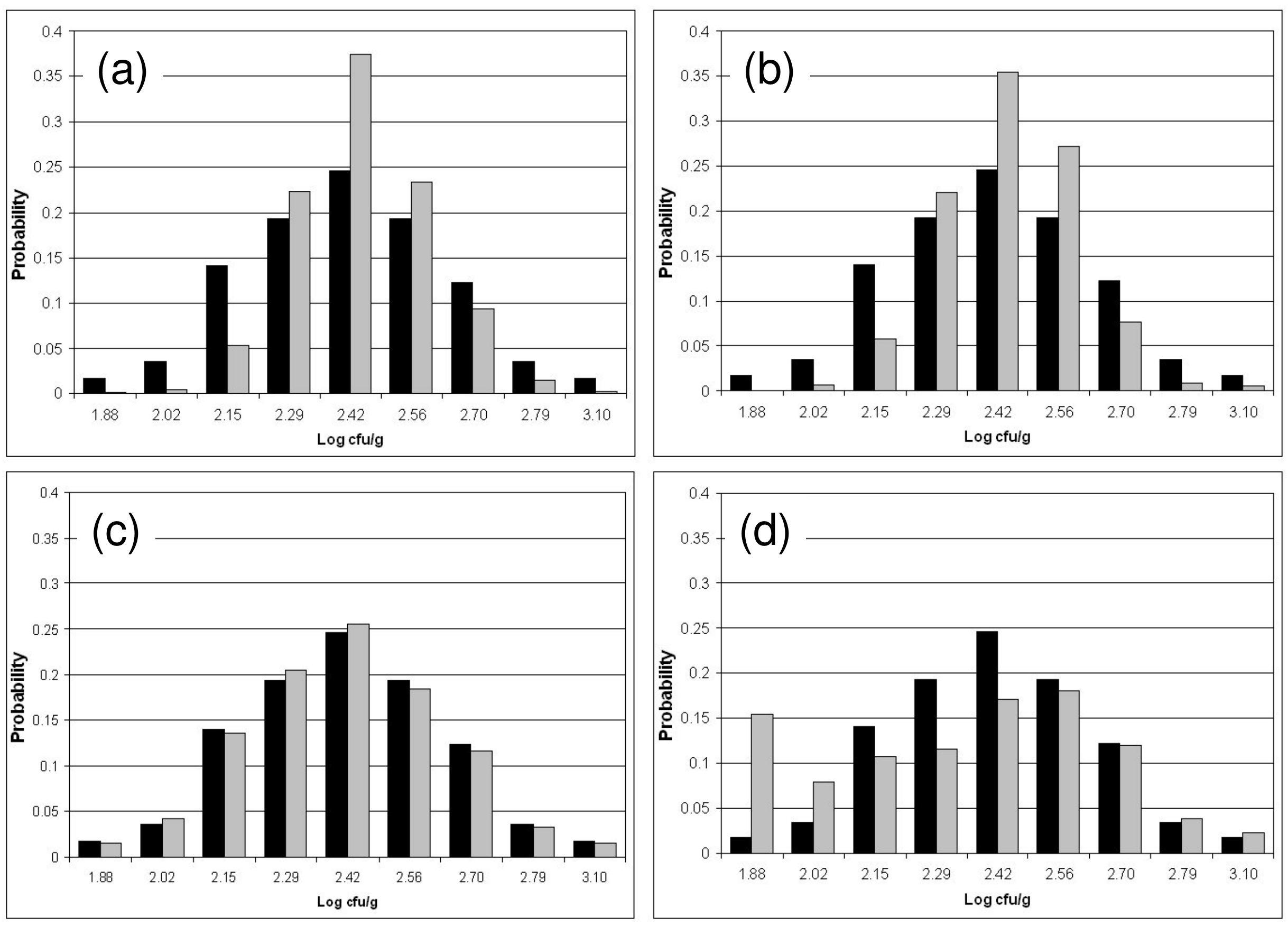}
\end{center} \vspace{-0.7cm}\caption{Theoretical distributions (white
bars) of the \emph{L. monocytogenes} concentration at 168 hours for
(a) $\sigma_T=0$, $\sigma_{pH}=0$, $\sigma_{aw}=0$, (b)
$\sigma_T=10^{-2}$, $\sigma_{pH}=5\cdot 10^{-4}$,
$\sigma_{aw}=10^{-5}$, (c) $\sigma_T=2\cdot 10^{-1}$,
$\sigma_{pH}=10^{-4}$, $\sigma_{aw}=1.5\cdot 10^{-4}$, (d)
$\sigma_T=5\cdot 10^{-1}$, $\sigma_{pH}=5\cdot 10^{-4}$,
$\sigma_{aw}=10^{-5}$. Black bars represent the corresponding
experimental distribution.\vspace{0.0cm}} \label{fig8}
\end{figure}
The results, obtained for suitable values of the interaction
parameters ($\beta_{Lmo/LAB}=0.656$, $\beta_{LAB/Lmo}=0$), are shown
in Fig.~\ref{fig7}. Here we note that the theoretical curves of
\emph{L. monocytogenes} (dashed black line) and LAB (full black
line) fit very well the corresponding experimental data (black
squares for \ emph{L. monocytogenes}, black circles for LAB). This
indicates that the interaction, present in the model, between the
two bacterial populations reproduces a feature of the real
biological system~\cite{Fia04,Val04b,Val08b}. In particular, we note
that the condition $\beta_{LAB/Lmo}=0$ implies the absence of any
direct effects of \emph{L. monocytogenes} on the dynamics of LAB. On
the other hand, the limiting effect of LAB on the growth of \emph{L.
monocytogenes}, obtained for a suitable positive value of the other
interaction parameter ($\beta_{Lmo/LAB}=0.665$), determines
conditions for the coexistence of the two populations, according to
empirical data~\cite{Cam93,Giu05,Tru89}. To analyze the role of the
random fluctuations on the dynamics ofthe system, we solve
Eqs.~(\ref{eq_Lmo})--(\ref{eqaw}) both in deterministic regime and
for three different values of the noise intensities $\sigma_T$,
$\sigma_{pH}$ and $\sigma_{aw}$. We obtain the theoretical
probability distributions of \emph{L. monocytogenes} concentration
at the end of the fermentation period (168 hours). Predicted
results, together with observed data, are shown in Fig.~\ref{fig8}.
Here, the histograms indicate that the best agreement between the
theoretical distribution (white bars) and experimental one (black
bars) is observed when the bacterial dynamics is obtained for values
of the noise intensities different from zero (stochastic dynamics),
and in particular for $\sigma_T=2\cdot 10^{-1}$,
$\sigma_{pH}=10^{-4}$, $\sigma_{aw}=1.5\cdot 10^{-4}$ (panel c).
This result accords with the complex nature of the system analyzed,
in which random fluctuations of environmental variables such as
temperature, pH and activity water, are present.

\section{Conclusions}\label{S:5}

In this paper we reviewed some recent results on effects of noise in
the dynamics of two competing populations, whose interaction depends
on a parameter $\beta$.

First we studied the dynamics of an ecosystem affected by two
sources of random fluctuations: a multiplicative noise and an
additive noise. The latter induces a coherent time behavior and
oscillating time series of the two species densities. Moreover, an
enhancement of the response of the system through stochastic
resonance phenomenon is observed as a function of the multiplicative
noise intensity. The model reproduces the dynamics of ecosystems
subject to both deterministic oscillating changes and random
modifications of environmental variables, such as variations of
temperature. This interplay between the deterministic and random
signals and the nonlinearity can determine a coherent response of
the ecosystem. Finally, additive noise causes also a delayed
extinction of one of the two populations. In particular, a
nonmonotonic behaviour of the mean extinction time, with a minimum,
is found as a function of the additive noise intensity.

Afterwards we presented links between theoretical modelling in
population dynamics and species distributions in two different real
ecosystems consisting of two populations.

First, results obtained from a discrete time evolution model were
compared with those recorded for the spatial distributions of two
pelagic fish populations, i.e. anchovies and sardines. The
comparison showed the presence of strong correlations between
theoretical and experimental distributions for both populations.
These findings, previously not published, represent the novelty of
the paper.

Then, we discussed a predictive microbiological model which allows
to describe microbial evolution in food products as a function of
environmental conditions. Our findings indicate that interspecific
bacterial interaction and environmental random fluctuations are
essential for a more precise and reliable prediction of the
bacterial dynamics.

The noise induced phenomena discussed in this paper can contribute
to understand population dynamics in ecosystems, which are complex
systems due to their intrinsic nonlinearity and continuous exchange
with the environment through deterministic and random
perturbations~\cite{Bjo01,Car02,Cir03,Gar02,LaB02,Spa02a,Spa02b,Zim99}.
In particular, the results presented in this paper highlight the
importance of including noise effects to model more effectively the
dynamics of two specific real ecosystems. Indeed our results could
contribute: (i) to reproduce the dynamics of fish populations and
predict the effects of global warming on marine ecosystems, in view
of devising fishing strategies which prevent the decline of marine
populations, such as sardines and anchovies, of paramount importance
for countries whose economy in strongly based on fishing activities;
(ii) to incorporate stochastic microbial predictive models into a
risk assessment process, and therefore to improve the precision of
the expected concentrations of a foodborne disease agent. This
aspect agrees to the new European approach to food risk assessment
and management.


\section*{Acknowledgements}
Authors acknowledge the financial support by Ministry of University,
Research and Education of Italian Government, Project
PON02\_00451\_3362121 "PESCATEC -- Sviluppo di una Pesca Siciliana
Sostenibile e Competitiva attraverso l'Innovazione Tecnologica", and
Project PON02\_00451\_3361909 "SHELF-LIFE -- Utilizzo integrato di
approcci tecnologici innovativi per migliorare la shelf-life e
preservare le proprietà nutrizionali di prodotti agroalimentari".


\end{document}